# Decoding Complex Compositions in Topologically Close-Packed Nanoplates of Magnesium Alloys: A High-Throughput Route to Stable Precipitates

Junyuan Bai[1], Xueyong Pang[1,2*], Gaowu Qin[1,3,4*]

[1]*Key Laboratory for Anisotropy and Texture of Materials (Ministry of Education), School of Materials Science and Engineering, Northeastern University, Shenyang 110819, China*
[2]*State Key Laboratory of Rolling and Automation, Northeastern University, Shenyang 110819, China*
[3]*Institute of Materials Intelligent Technology, Liaoning Academy of Materials, Shenyang 110004, China*
[4]*Research Center for Metal Wires, Northeastern University, Shenyang 110819 China*

**Abstract**:

Coherent topologically close-packed (TCP) nanoplates in magnesium alloys can effectively improve strength and creep resistance. However, the formation mechanisms of several metastable TCP nanoplates remain unclear, and the traditional trial-and-error methods hinder the rapid discovery of novel TCP precipitate-strengthened Mg alloys. In this study, using density functional theory calculations to construct convex hull diagrams and evaluate phase stability, our results reveal that the metastable $\beta_2^{'}$ nanoplates in Mg-Zn alloys adopt the Mg(Mg, Zn)$_2$ composition rather than pure MgZn$_2$, resolving the longstanding puzzle of why stable MgZn$_2$ manifests as metastable nanoplates during Mg-Zn alloy aging. By integrating the thermodynamic and kinetic conditions for TCP precipitation, we developed a two-step high-throughput screening strategy, identifying 43 previously unreported TCP nanoplates in a series of Mg alloys. These findings highlight the critical need for precise compositional characterization of nanoprecipitates and establish a theoretical framework for designing creep-resistant Mg alloys containing TCP nanoplates.

Corresponding author.

Email: pangxueyong@mail.neu.edu.cn (X.Y. Pang)
qingw@smm.neu.edu.cn (G. W. Qin)

Magnesium alloys, as the lightest structural metallic materials, have attracted considerable research attention for their applications in aerospace and automotive industries, wherein weight reduction is crucial for lowering energy consumption and $CO_2$ emissions. In Mg-alloy development, precipitation hardening is often recognized as an effective strengthening strategy, achieved by forming dense, coherent nanoprecipitates during aging treatments [1]. Among these nanoprecipitates, coherent topologically close-packed (TCP) nanoplates along $\{0001\}_{hcp}$ basal planes are particularly significant, as they not only provide exceptional strength enhancement but also improve creep resistance, making them ideal for elevated-temperature applications [1–4]. These TCP nanoplates have been identified across various alloy systems, including stable phases like $Mg_2Ca$ and $Al_2Ca$ phases in Mg-Ca-(Al) alloys[5–7], metastable $\beta_2'$-$MgZn_2$ phase in Mg-Zn alloys[8,9], and single unit-cell height γ″ metastable phases in Mg-RE (rare earth)-(Zn, Ag) series alloys[4,10,11]. Given the outstanding performance of TCP nanoplates in Mg alloys, the systematic identification of alloy systems capable of forming TCP nanoplates has become a crucial priority in Mg alloy development.

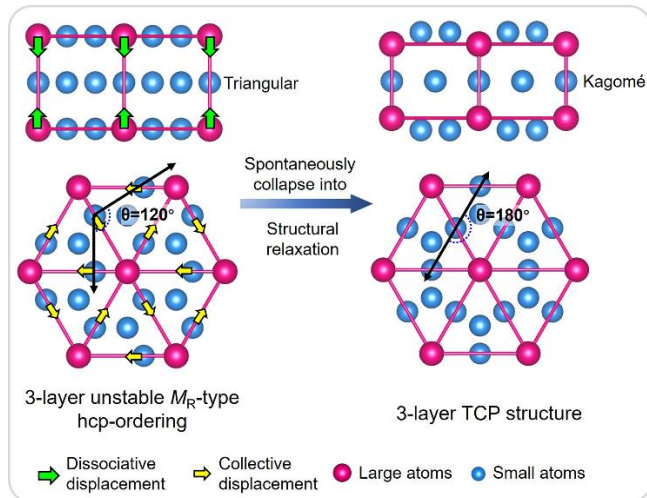

**Figure 1.** Schematic illustration of the spontaneous transformation of unstable 3-layer $M_R$-type hcp-ordering into a 3-layer TCP structure upon structural relaxation[12].

However, traditional trial-and-error methods are costly and inefficient for rapid alloy development. Although experimental phase diagrams and density functional theory (DFT)-calculated convex hull diagrams provide thermodynamic insights into equilibrium or metastable phases[13], they cannot determine whether a TCP phase can precipitate from the parent matrix (a process intricately associated with kinetics) and what its resulting morphology will be. Our recent study[12] reveals that the full structural evolution pathway of TCP nanoplates within the hcp-Mg matrix is governed by the formation of 3-layer $\{0001\}_{hcp}$-oriented unstable solute clusters (designated as $M_R$-type hcp-ordering, Fig. 1), involving only shuffle-based displacements. During aging, under thermal fluctuations, once solute clusters evolve into the 3-layer $M_R$-type structures at a specific moment, they will spontaneously collapse into a TCP structure. Remarkably, this structural instability of specific solute clusters can be identified by structural relaxation in DFT calculations, establishing a kinetic criterion for TCP formation: if the 3-layer $M_R$-type hcp-ordering fails to induce hcp→TCP transitions upon structural relaxation, coherent TCP nanoplates cannot form kinetically.

Meanwhile, understanding the formation mechanism of TCP nanoplates from a thermodynamic perspective is also crucial. Here, we first use several common TCP nanoplates to elucidate the thermodynamic conditions for the precipitation of stable and metastable TCP nanoplates from typical dilute Mg alloys and explain the physical mechanism behind complex experimental phenomena. For stable/equilibrium phase precipitation, taking the $Mg_2Ca$ and $Al_2Ca$ phases in Mg-Ca-(Al) alloys as examples, we reconstructed the 0 K Mg-Ca and Mg-Al-Ca convex hull diagrams (see Fig. 2) using structural data from the Open Quantum Materials Database (OQMD)[14,15]. Convex hull analysis is a powerful method for evaluating phase stability. A phase is thermodynamically stable if its formation energy at a given composition is lower than that of all other phases of the same composition and any linear combination of energies of other phases in the phase space. The collection of all stable phases in a given phase space constitutes the convex hull, while phases located above the convex hull are thermodynamically unstable. The parameter $E_{hull}$ measures the distance from the convex hull line and quantifies thermodynamic stability, with stable phases residing exactly on the hull ($E_{hull}$=0.0 eV/atom) and metastable/unstable phases located above it ($E_{hull}$>0.0 eV/atom). According to convex hull theory[16], in binary or ternary systems, unstable phases with compositions between stable phases will decompose into their neighboring stable phases, while those above a stable phase with identical composition will transform into that stable phase.

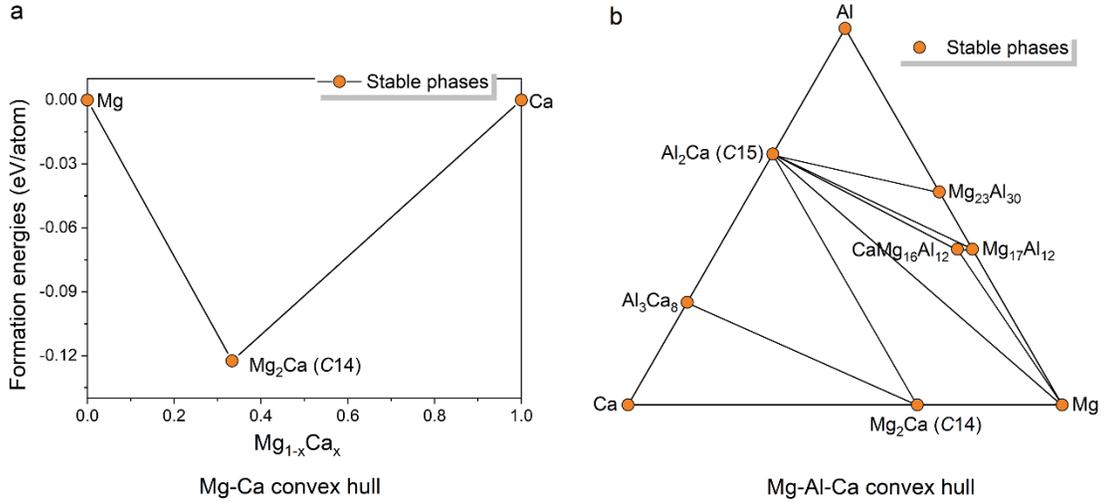

**Figure 2**. DFT-calculated zero-temperature Mg-Ca and Mg-Al-Ca convex hull diagrams.

Figure 2 shows that both Mg$_2$Ca with a $C$14 structure and Al$_2$Ca with a $C$15 structure lie on the convex hull, indicating that these two Laves structures are stable phases. Additionally, they also form a tie-line with Mg. A tie-line represents a straight line connecting two stable phases on a convex hull diagram, meaning that two phases can coexist. Therefore, Mg-Ca-(Al) alloys of suitable compositions thermodynamically favor the precipitation of stable Mg$_2$Ca and Al$_2$Ca phases during aging. In convex hull diagrams, precipitation of stable TCP phase from the parent Mg matrix not only demands that phase displays $E_{hull}$=0.0 eV/atom but also forms a tie-line with Mg. In contrast, the thermodynamic precipitation mechanism of metastable TCP nanoplates is more intricate. Here, we first aim to explain why the originally stable MgZn$_2$ phase in the Mg-Zn system occurs in the form of metastable $\beta_2^{'}$-MgZn$_2$ nanoplates during the aging of Mg-Zn alloys.

The experimental Mg-Zn phase diagram[17] and the Mg-Zn convex hull diagram (see Fig. 3b) both indicate that the MgZn$_2$-$C$14 structure is a stable phase. According to the established thermodynamic conditions for stable-phase precipitation, this MgZn$_2$-$C$14 phase should form a tie-line with Mg. Instead, there are Mg$_{21}$Zn$_{25}$ and Mg$_4$Zn$_7$ stable phases situated between MgZn$_2$-$C$14 and Mg, preventing the precipitation of a stable MgZn$_2$-$C$14 phase from Mg-Zn alloys. However, recent experiments [9,18] commonly suggest the precipitation sequence in Mg-Zn alloys as supersaturated solid solution (S.S.S) → GP zones → $\beta_1^{'}$→ $\beta_2^{'}$ (MgZn$_2$ nanoplates) →$\beta$-Mg$_{21}$Zn$_{25}$, with $\beta_2^{'}$-MgZn$_2$ nanoplates clearly identified as metastable intermediates. This naturally raises the question of why the originally stable MgZn$_2$ phase forms during the aging of Mg-Zn alloys and exists in a metastable form.

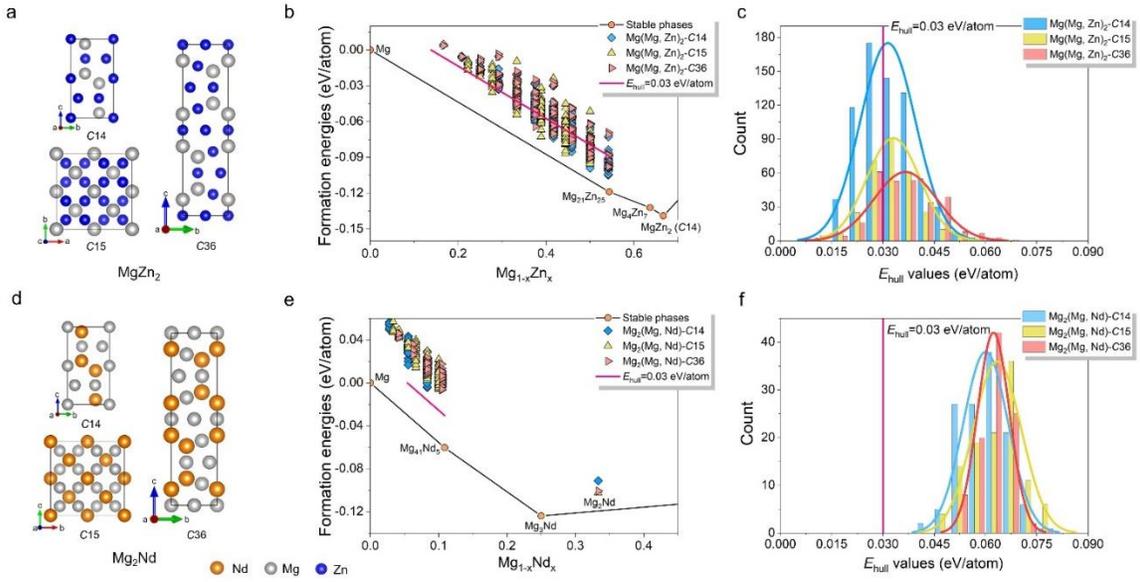

**Figure 3.** a. Atomic configurations of $C$14, $C$15, and $C$36-structured MgZn$_2$ phase. b. DFT-calculated Mg-Zn convex hull diagram. c. The distribution statistics of $E_{hull}$ values for $C$14, $C$15, and $C$36-based Mg(Zn, Mg)$_2$ structures. d. Atomic configurations of $C$14, $C$15, and $C$36-structured Mg$_2$Nd phase. e. DFT-calculated Mg-Nd convex hull diagram. f. The distribution statistics of $E_{hull}$ values for $C$14, $C$15, and $C$36-based Mg$_2$(Mg, Nd) structures.

In reality, nanosized precipitates often have non-stoichiometric compositions with excess matrix elements, as evidenced in systems like the Al$_2$Ca phase in Mg-Al-Ca alloys[19], Al$_3$Sc in Al-Sc alloys[20], and $\eta$-MgZn$_2$ phases in Al-Mg-Zn alloys[21,22]. Our recent research[12] demonstrates that TCP precipitation in the Mg matrix follows a nonclassical nucleation mechanism, where hcp→TCP transitions mainly rely on the distribution of larger atoms in the nucleus rather than smaller ones. For example, during the formation of the MgZn$_2$ phase (see Fig. S5), an inhomogeneous nucleus with Mg occupying anti-sites (Mg in Zn positions) can still trigger hcp→TCP transitions. Given the nonclassical nucleation characteristics of TCP formations and the constraints of precipitation dynamics, realistic TCP nanoplates readily retain excess matrix Mg. In this context, the metastable $\beta_2^{'}$-MgZn$_2$ nanoplates may also incorporate excess Mg, resulting in Mg(Zn, Mg)$_2$ stoichiometry. To evaluate the thermodynamic feasibility of Mg(Zn, Mg)$_2$ precipitation, we mixed varying amounts of Mg into the Zn positions within the three basic $C$14, $C$15, and $C$36 Laves structures (Fig. 3a). The formation energies of these Mg(Zn,

Mg)$_2$ structures were plotted on the Mg-Zn convex hull diagram to assess their phase stability. The Method section in Supplementary Information details the generation of various Mg(Zn, Mg)$_2$ structures, and the distribution statistics of their $E_{\text{hull}}$ values are shown in Fig. 3c.

Figure 3b shows that the formation energies of all Mg(Zn, Mg)$_2$ structures lie above the convex hull line, indicating their metastable nature. $E_{\text{hull}}$ is typically utilized as an indicator of phase stability to predict the formability of a material, where lower $E_{\text{hull}}$ values suggest higher thermodynamic feasibility. Although the exact boundary distinguishing formable and non-formable materials remains undefined, a general threshold of $E_{\text{hull}} \leqslant 0.03$ eV/atom is often considered a rule of thumb for the possible formation of metastable phases[16,23]. In Fig. 3c, a large proportion of Mg(Zn, Mg)$_2$ structures (where excess Mg atoms occupy Kagomé layers), including $C$14, $C$15, and $C$36 types, fall below this threshold. This distribution proves the thermodynamic feasibility of metastable Mg(Zn, Mg)$_2$ precipitation and predicts that $C$14, $C$15, and $C$36 structures could all present in $\beta_2'$ nanoplates. Experimental observations[24] have reported $\beta_2'$ nanoplates with both $C$14 and $C$15 structures. Therefore, we rectify that the stoichiometry of metastable $\beta_2'$ nanoplates should be Mg(Zn, Mg)$_2$ rather than pure MgZn$_2$. Moreover, careful characterization of compositions is crucial for both stable and metastable nanoprecipitates, particularly in identifying the incorporation of matrix elements, although this is always an experimental challenge.

This study further extends to elucidate the thermodynamic precipitation mechanism of metastable Mg$_2$Nd Laves nanoplates, which exhibit unique system-dependent formation behavior. A striking observation is that these metastable Mg$_2$Nd exclusively appear in the Mg-Nd-Ca ternary system[25], while remaining absent in binary Mg-Nd alloys. In Mg-Nd binary alloys, the reported final equilibrium precipitate is Mg$_{41}$Nd$_5$[18,26], consistent with the Mg-Nd convex hull diagram (see Fig. 3e), where Mg$_{41}$Nd$_5$ is the stable phase forming a tie-line with Mg. Since the metastable Mg$_2$Nd lies outside the Mg$_{41}$Nd$_5$-Mg region, its precipitation from Mg-Nd alloys is thermodynamically impossible. When further considering whether nonclassical nucleation behavior (with Mg occupying Nd positions) could facilitate the precipitation of metastable Mg$_2$(Nd, Mg) phases, our evaluation of phase stability for $C14$, $C15$, and $C36$-derived $Mg_2(Nd, Mg)$ structures (Fig. 3e, f) shows that most have positive formation energies and $E_{hull}$ values well above 0.03 eV/atom. Thus, the precipitation of metastable $Mg_2(Nd, Mg)$ nanoplates from Mg-Nd alloys is thermodynamically unfeasible.

In Mg-Nd-Ca ternary alloys, we found that the formation of $Mg_2(Ca, Nd, Mg)$ metastable structures by incorporating Nd and excess Mg into $Mg_2Ca$ nanoplates is thermodynamically feasible. The Mg-Nd-Ca convex hull diagram (Fig. 4a) shows an Mg-$Mg_2Ca$-$Mg_{41}Nd_5$ triangle on the Mg-rich side. When introducing Nd and Mg into Ca positions and generating $C14$, $C15$, and $C36$-based $Mg_2(Nd, Mg)$ structures, where their compositions require to lie within the Mg-$Mg_2Ca$-$Mg_{41}Nd_5$ triangle, the $E_{hull}$ values of generated $Mg_2(Nd, Mg)$ structures have a large amount of portion fall below 0.03 eV/atom. This demonstrates the thermodynamic feasibility of metastable $Mg_2(Nd, Mg)$ precipitation. Thus, under the structural templating of the stable $Mg_2Ca$ phase, the formation of metastable TCP nanoplates with complex compositions is accessible. Experimentally, when using high-angle annular dark-field scanning transmission electron microscopy (HAADF-STEM) to characterize nanoprecipitates, the high $Z$-contrast of Nd easily causes researchers to focus solely on Nd, overlooking lower $Z$-contrast elements like Ca and Mg. We once again highlight the critical need for accurate compositional determination of nanoprecipitates.

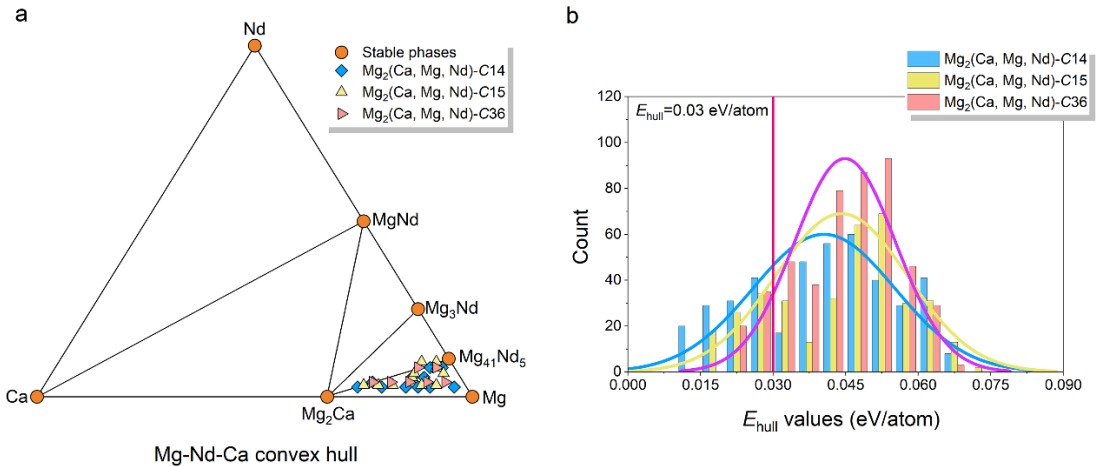

**Figure 4**. a. DFT-calculated zero temperature Mg-Nd-Ca convex hull diagram. b. The distribution statistics of $E_{hull}$ values for $C14$, $C15$, and $C36$-based $Mg_2(Ca, Mg, Nd)$ structures.

Although metastable TCP nanoplates in Mg alloys may offer temporary creep-resistance improvements, stable TCP nanoplates are more desirable for developing Mg alloys with superior long-term high-temperature creep resistance. This study devises a two-step screening strategy (Fig. 5) to identify Mg alloy systems capable of forming stable TCP nanoplates. By combining thermodynamic and kinetic conditions for TCP precipitation, this strategy utilizes DFT-calculated convex hull diagrams from the OQMD[14,15] to search for stable TCP phases forming a tie-line with Mg, followed by evaluating the structural stability of corresponding $M_R$-type $X$-ordering to determine the kinetic feasibility of TCP formation. TCP phases in Mg alloys, including Laves ($AB_2$-type) and Laves-like ($AB_5$, $AB_3$, and $A_2B_7$-types) phases, can be categorized into three groups: Mg-containing binary phases (e.g., $Mg_2Ca$) in binary alloys, Mg-free binary (e.g., $Al_2Ca$) and Mg-containing ternary phases (e.g., $LaMg_2Cu_9$) in ternary alloys. A total of 64 Mg-$X$ binary and 2022 Mg-$X$-$Y$ ternary convex hull diagrams from the OQMD were employed in this screening process.

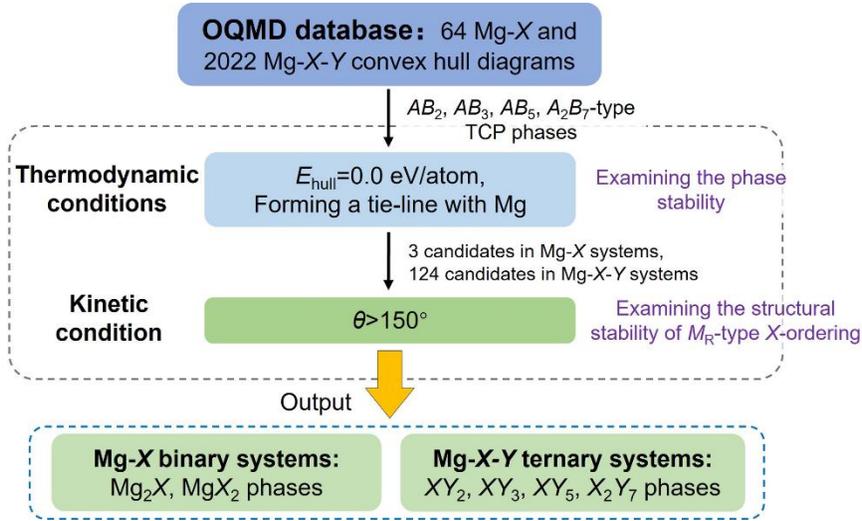

**Figure 5.** A two-step screening strategy to identify effective alloying elements in Mg-$X$ binary and Mg-$X$-$Y$ ternary alloys that can form coherent stable TCP nanoplates.

During the second-step screening, since the nonclassical nucleation nature in hcp→TCP transformations dictates that Mg-free TCP phase nucleation predominantly depends on the distribution of large atoms instead of small ones[27], only $M_R$-type $X$-ordering needs to be examined in Mg-$X$-$Y$ alloys (where $X$ denotes large atoms and $Y$ small atoms), obviating the need for extensive examinations of $M_R$-type $X$-$Y$-orderings across 124 candidates. Given that hcp→TCP transitions involve a lattice transformation from triangular to kagomé structures (which changes the θ value from 120° to 180°[27]), as shown in Fig. 1, a larger change in θ value within our model correlates with a higher tendency for TCP formation. Hence, we adopted θ as a structural indicator to examine the structural stability of $M_R$-type $X$-orderings, setting a threshold of θ>150° based on the observed clustering distribution of θ values across various Mg-X combinations (see Fig. S4). For this screening step, we used 162-atom models (Fig. S1 (a-b)) with different $M_R$-type $X$-ordering sizes for structural relaxations. Further details on DFT calculations and models are provided in the Supplementary Information.

**Table 1.** High-throughput screening results of stable TCP nanoplates in Mg-$X$ binary and Mg-$X$-$Y$ ternary alloys.

| No. | Mg-$X$ binary systems | Structural type | No. | $X$ | Mg-$X$-$Y$ systems | Structural type |
|---|---|---|---|---|---|---|
| 1 | Mg$_2$Ca† | $C$14 | 21 | | HoFe$_2$ | $C$15 |
| 2 | Mg$_2$Yb | $C$14 | 22 | | HoFe$_3$ | $AB_3$ |
| 3 | MgCo$_2$ | $C$36 | 23 | Ho | HoOs$_2$ | $C$14 |
| No. | $X$ | Mg-$X$-$Y$ systems | Structural type | 24 | | HoRe$_2$ | $C$14 |
| 1 | Ca | CaAl$_2$† | $C$15 | 25 | | HoTc$_2$ | $C$14 |
| 2 | Ce | CeAl$_2$ | $C$15 | 26 | K | KNa$_2$ | $C$14 |
| 3 | Cs | CsK$_2$ | $C$15 | 27 | La | LaAl$_2$ | $C$15 |
| 4 | | CsNa$_2$ | $C$14 | 28 | Nd | NdAl$_2$ | $C$15 |
| 5 | | DyAl$_2$ | $C$15 | 29 | Pm | PmAl$_2$ | $C$15 |
| 6 | Dy | DyOs2 | $C$14 | 30 | Pr | PrAl$_2$ | $C$15 |
| 7 | | DyRe$_2$ | $C$14 | 31 | Sm | SmAl$_2$ | $C$15 |
| 8 | | DyTc$_2$ | $C$14 | 32 | | SmOs$_2$ | $C$14 |
| 9 | | ErAl$_2$ | $C$15 | 33 | Tb | TbAl$_2$ | $C$15 |
| 10 | | ErCo$_2$ | $C$15 | 34 | | TbOs$_2$ | $C$14 |
| 11 | | ErFe$_2$ | $C$15 | 35 | | Th$_2$Fe$_7$ | $A_2B_7$ |
| 12 | Er | ErFe$_3$ | $AB_3$ | 36 | | ThFe$_5$ | $AB_5$ |
| 13 | | ErOs$_2$ | $C$14 | 37 | Th | ThOs$_2$ | $C$14 |
| 14 | | ErRe$_2$ | $C$14 | 38 | | ThRe$_2$ | $C$14 |
| 15 | | ErTc$_2$ | $C$14 | 39 | | ThTc$_2$ | $C$14 |

| 16 | Eu | EuAl$_2$ | C15 | 40 |  | YAl$_2$† | C15 |
| 17 |  | GdAl$_2$† | C15 | 41 | Y | YOs$_2$ | C14 |
| 18 | Gd | GdOs$_2$ | C14 | 42 |  | YRe$_2$ | C14 |
| 19 |  | GdRe$_2$ | C14 | 43 |  | YTc$_2$ | C14 |
| 20 | Ho | HoAl$_2$ | C15 | 44 | Yb | YbAl$_2$ | C15 |

† represents the TCP phase that has been observed in experiments.

Table 1 summarizes the high-throughput screening results for stable TCP phases in Mg-$X$ and Mg-$X$-$Y$ alloys. In binary systems, Mg$_2$Ca, Mg$_2$Yb, and MgCo$_2$ were identified as potential precipitates, with Mg$_2$Ca nanoplates have been experimentally confirmed. For ternary Mg-$X$-$Y$ alloys, 44 TCP phases, including Al$_2$Ca [5], Al$_2$Y [28], and Al$_2$Gd [29] (Laves precipitates widely observed in previous research), were found, demonstrating the reliability of our screening approach. As TCP nanoplates are known to enhance the strength and creep resistance of Mg alloys, the newly discovered TCP phases could expand the chemical space for alloy design. Table 1 further suggests that REAl$_2$ Laves phases may widely precipitate in Mg-RE-Al series alloys. In these alloys [30], TCP nanoplates typically form at lower aging temperatures (~200°C), while at higher temperatures (~500°C), long-period stacking ordered (LPSO) phases become more common.

A distinct class of single-unit-cell-thick metastable TCP phases, termed γ″ phases, are often observed in Mg-RE-Zn (Ag) alloys. Owing to their unique ability to maintain constant thickness during aging, these γ″ phases have received widespread attention over the past decade[4,10,18]. Regarding their thermodynamic origins, Zhao et al.[31] recently suggest that γ″ phases in Mg-Nd-Ag alloys act as metastable intermediates for the equilibrium non-TCP NdAgMg$_{11}$ phase, based on HAADF-STEM observations and atomic-scale energy-dispersive X-ray spectroscopy. This reveals a unique precipitation pathway where non-TCP equilibrium phases use TCP-structured intermediates to mediate their precipitation. Applying the two-step screening strategy, we extended our search to potential stable TCP phases in hcp-structured Ti and Zr alloys in the Supplementary Information. Our results indicate that only the TiCr$_2$ and ZrMn$_2$ Laves phases (see Table S1) may precipitate from Ti-Cr and Zr-Mn binary alloys. This is mainly because, under the formation of unstable $M_R$-type hcp-ordering, all large elements in the Ti (Zr) matrix cannot interact with the surrounding matrix to induce hcp→TCP transitions (see Fig. S4a), unlike those in the soft Mg matrix. Meanwhile, we emphasize that these screening results offer only preliminary guidance for future experimental explorations, as many intricate factors, such as solid solubility of relevant adding elements, alloy composition, thermal processing history, etc., also significantly affect the precipitation of TCP plates.

In this study, using DFT calculations to construct convex hull diagrams and assess phase stability, we rectified that the metastable $\beta_2'$ nanoplates in Mg-Zn alloys are Mg(Mg, Zn)$_2$ rather than pure MgZn$_2$. We also clarified that the stoichiometry of the experimentally observed metastable Mg$_2$Nd phase in Mg-Nd-Ca alloys is Mg$_2$(Nd, Mg, Ca) and explained why this phase does not form in binary Mg-Nd alloys. This highlights the importance of accurately determining the composition of nanoprecipitates in experiments. Furthermore, by integrating the thermodynamic and kinetic conditions for TCP precipitation, we developed a two-step high-throughput screening strategy to identify stable TCP nanoplates in Mg alloys. The newly discovered TCP phases may facilitate the design of high-strength and creep-resistant Mg alloys. These findings enhance our atomic-level understanding of precipitation behavior in various hcp-based alloys and lay a foundation for designing advanced materials incorporating coherent TCP nanoplates.

**Acknowledgments**

This research is supported by the National Key Research and Development Program of China (2023YFB3710902), Fundamental Research Funds for the Central Universities of China (N2102011, N2007011, N160208001), and National 111 Project (B20029).

**Author contributions**

G. Qin and J. Bai conceived the original idea and designed the work. J. Bai conducted the simulations with help from X. Pang. J. Bai wrote the paper. G. Qin supervised the project and revised the manuscript.

**Declaration of Competing Interests**: The authors declare that they have no known competing financial interests or personal relationships that could have appeared to influence the work reported in this paper.

**Data availability**

The data that support the findings of this study are available from the corresponding author upon reasonable request.

**Code availability**

The code in this study is available from the authors upon reasonable request.

# Supplementary Information for

# Decoding Complex Compositions in Topologically Close-Packed Nanoplates of Magnesium Alloys: A High-Throughput Route to Stable Precipitates

Junyuan Bai[1], Xueyong Pang[1,2*], Gaowu Qin[1,3,4*]

[1]*Key Laboratory for Anisotropy and Texture of Materials (Ministry of Education), School of Materials Science and Engineering, Northeastern University, Shenyang 110819, China*
[2]*State Key Laboratory of Rolling and Automation, Northeastern University, Shenyang 110819, China*
[3]*Institute of Materials Intelligent Technology, Liaoning Academy of Materials, Shenyang 110004, China*
[4]*Research Center for Metal Wires, Northeastern University, Shenyang 110819 China*

Corresponding author.

Email: pangxueyong@mail.neu.edu.cn (X.Y. Pang)
qingw@smm.neu.edu.cn (G. W. Qin)

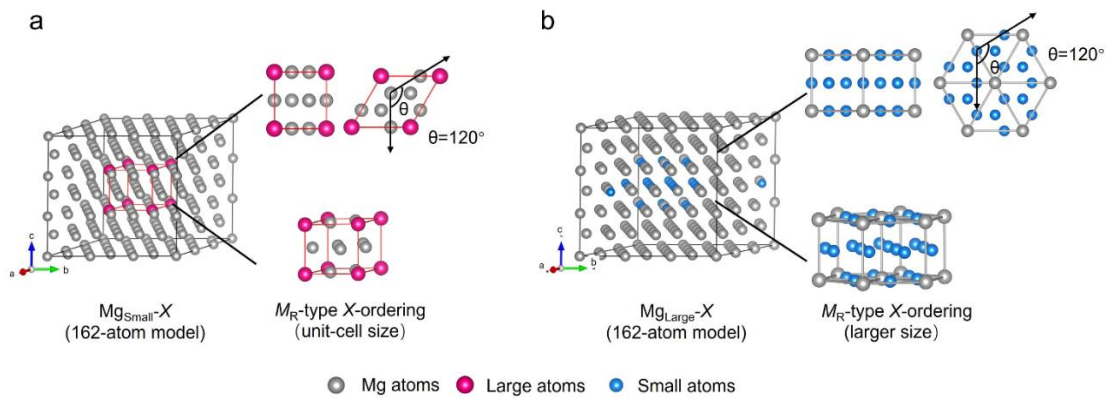

**Figure S1. a-b.** Atomic structural models in the situations of Mg$_{Small}$-X and Mg$_{Large}$-X. The Mg, large, and small atoms are marked with grey, dark-pink, and light-blue balls, respectively.

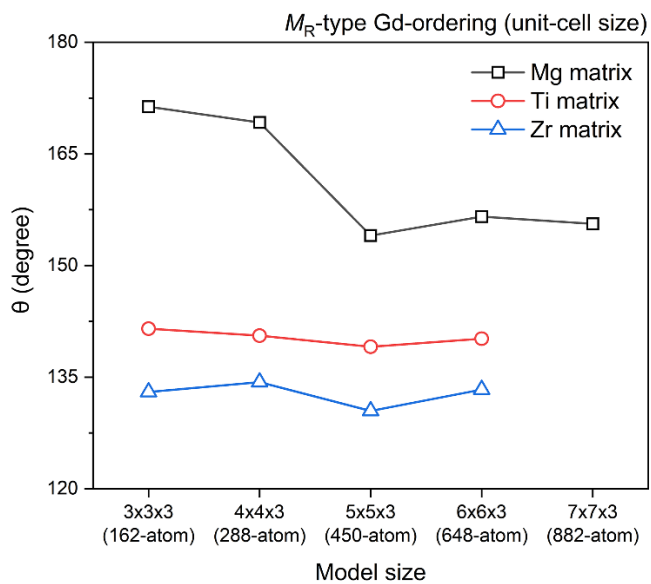

**Figure S2.** Convergence tests of θ value for model size across different matrices.

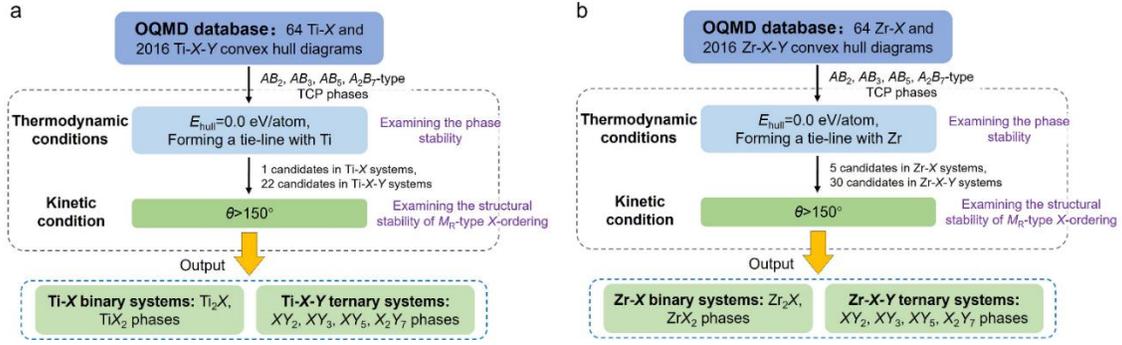

**Figure S3.** A two-step screening strategy to identify effective alloying elements in (Ti, Zr)-*X* binary and (Ti, Zr)-*X*-*Y* ternary alloys that can form coherent stable TCP nanoplates.

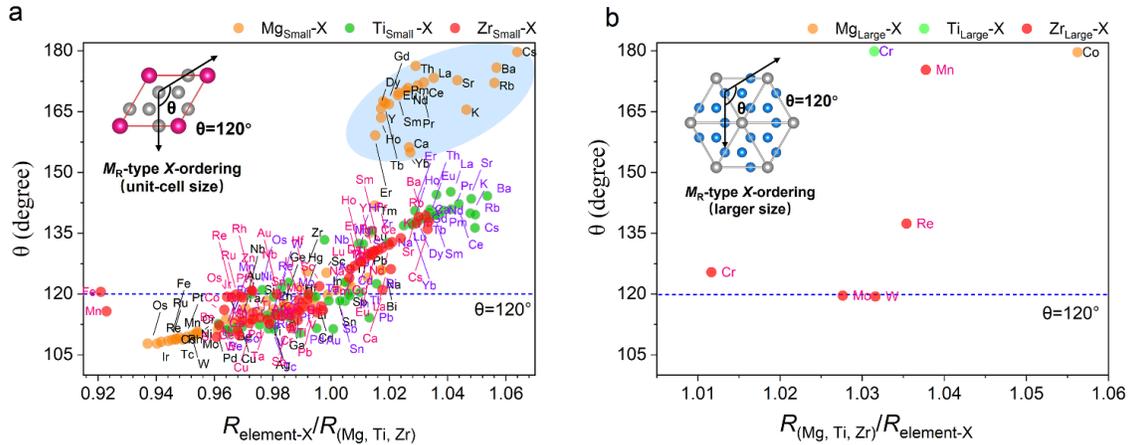

**Figure S4**. **a**. The variation of θ value of relaxed $M_R$-type *X*-ordering with calculated radius ratios ($R_{\text{element-X}}/R_{\text{(Mg, Ti, Zr)}}$) for (Mg, Ti, Zr)$_{\text{Small}}$-X scenarios. A unit-cell structure of $M_R$-type hcp-ordering is shown in the inset of the panel, and the pink and grey balls represent the large and small atoms, respectively. **b**. The variation of θ value of relaxed $M_R$-type *X*-ordering with calculated radius ratios ($R_{\text{(Mg, Ti, Zr)}}/R_{\text{element-X}}$) for (Mg, Ti, Zr)$_{\text{Large}}$-X scenarios. A large-sized structure of $M_R$-type *X*-ordering is shown in the inset of the panel, and the grey and blue balls represent the large and small atoms, respectively.

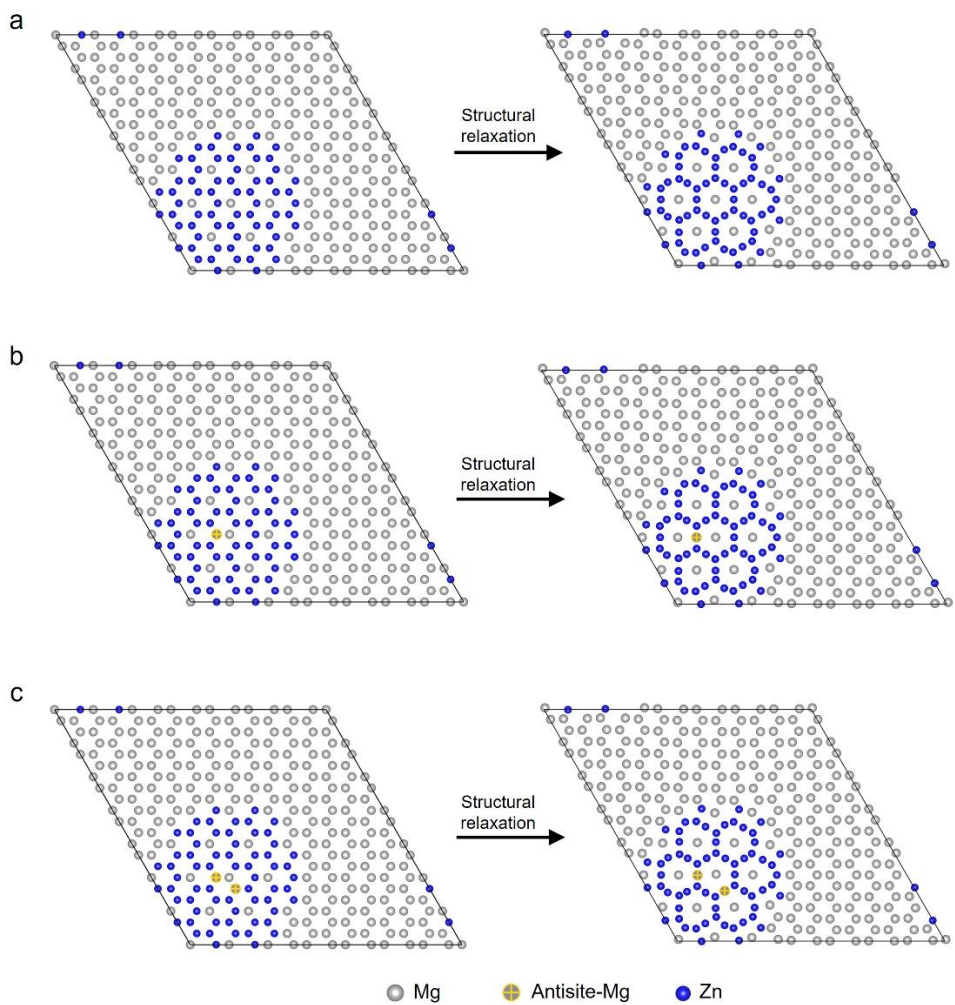

**Figure S5.** Schematic illustration of nonclassical nucleation behavior of MgZn$_2$ nucleation in the Mg matrix, depicting nuclei with (**a**) non antisite Mg atoms, (**b**) one antisite Mg atom, and (**c**) two antisite Mg atoms. An 882-atom model was employed for DFT calculations.

**Table S1**. High-throughput screening results of the stable TCP nanoplates in Ti-*X* and Zr-*X* binary alloys.

| No. | Ti-*X* binary systems | Structural type | No. | Zr-*X* binary systems | Structural type |
|---|---|---|---|---|---|
| 1 | TiCr$_2$ | *C*15 | 1 | ZrMn$_2$ | *C*15 |

**Methods**

DFT calculations were conducted using the Vienna *ab initio* simulation package (VASP)[1,2] employing Blochl's projector augmented wave (PAW) potential method[3]. The exchange-correlation energy functional was described with the generalized gradient approximation (GGA) as parameterized by Perdew-Burke-Ernzerhof (PBE)[4]. The frozen core pseudopotentials were used for RE elements, as these pseudopotentials have been shown to replicate the thermodynamics and elastic properties of rare-earth intermetallics [5,6]. A 520 eV plane wave cutoff was adopted, with convergence criteria for energy and atomic force set as $10^{-6}$ eV and $10^{-2}$ eV/Å, respectively. Partial occupancies were determined using the first-order Methfessel-Paxton method with a smearing width of 0.2 eV[7]. Relaxations of atomic coordinates and optimizations of the shape and size of the model were adopted for all calculations. A Γ-centered *k*-point mesh of 18×18×10 was adopted for the HCP primitive cell, with other supercells appropriately scaled to maintain a constant *k*-point density. Large numbers of symmetrically distinct Mg(Zn, Mg)$_2$, Mg$_2$(Nd, Mg), and Mg$_2$(Nd, Ca, Mg) structures based on *C*14, *C*15, and *C*36 types were generated with the Clusters Approach to Statistical Mechanics (CASM) code[8].

The unit-cell structure of $M_R$-type hcp-ordering was constructed based on their *a*-vector, *b*-vector, and *c*-vector parallel to the $[10\bar{1}0]_{hcp}$, $[01\bar{1}0]_{hcp}$ and $[0001]_{hcp}$ directions, respectively. During screening, we chose the θ value as a structural indicator to determine whether there were hcp→TCP transitions. In Fig. S2, we tested θ variation across different model sizes for three types of matrices. The results indicate that while smaller models, like the 162-atom (3×3×3) model (Fig. S1), amplify the θ value for the Mg matrix scenario due to periodic interactions, our calculations (Fig. S4a) based on this 162-atom model still display a clustered data distribution that can effectively identify candidate elements from a large pool of elements. Hence, to balance computational accuracy and cost, we employed this 162-atom (3×3×3) model containing $M_R$-type *X*-ordering with one unit-cell size or larger-sized (i.e., clusters in a 3D environment) for high-throughput screenings of hcp-Mg, Ti, and Zr alloys.

Models containing $M_R$-type $X$-ordering with a single unit-cell size (Fig. S1a, $Mg_{Small}$-X situations) were employed to examine Mg-free binary phases such as the $Al_2Ca$ phase, while the models containing the $M_R$-type $X$-ordering with larger size (Fig. S1b, $Mg_{Large}$-X situations) were used to examine the Mg-containing phases where Mg acts as the large atoms, e.g., $MgCo_2$ phases. The same rule was also applied to hcp-based Ti and Zr matrices. The calculated structural stability results of these $M_R$-type $X$-ordering across $Mg_{Small}$-X and $Mg_{Large}$-X situations within the Mg matrix, as well as in the matrices of hcp-based Ti, and Zr, are shown in Fig. S4. The atomic radii of different elements $R_{element}$ in various hcp-matrix were calculated by measuring the average nearest distance of neighboring matrix atoms in a 54-atom model.

The formation energy $E_f$ of a configuration $A_xB_y$, for instance, was calculated relative to the zero kelvin total energies of pure elements A and B as follows:

$$E_f = \frac{E(A_xB_y) - N_x E_A - N_y E_B}{N_x + N_y} \tag{S-1}$$

Where $E(A_xB_y)$ denotes the total energies of a configuration, and $E_A$ and $E_B$ are the total energies per atom for pure elements $A$ and $B$, respectively. $N_x$ and $N_y$ represent the number of elements $A$ and $B$ in the configuration. In this way, the formation energy of a configuration with ternary or more constituents can be obtained. The zero temperature convex hull diagrams were constructed by using the *pymatgen* code [9].

**Supplementary References:**